\documentstyle[prc,aps,preprint,tighten,graphicx]{revtex}

\def\hyp#1#2{${}^{#1}_\Lambda$#2}
\def\gngp{$\Gamma_n/\Gamma_p$\ }             
\begin{document}
\bibliographystyle{alpha}
\date{\today}

\title{Final State Interactions in Hypernuclear Decay}
\author{A. Parre\~no\footnote{Present address: Departament d'Estructura i 
Constituents de la Mat\`eria, Universitat de Barcelona, Diagonal 647,
E-08028 Barcelona, Spain}}

\address{Institute for Nuclear Theory, University of Washington,
98195-1550 Seattle, WA, USA} 

\author{A. Ramos}

\address{Departament d'Estructura i
Constituents de la Mat\`eria, Universitat de Barcelona, Diagonal 647,
E-08028 Barcelona, Spain}       

\maketitle

\begin{abstract}

We present an update of the One-Meson-Exchange (OME) results for the
weak decay of s- and p-shell hypernuclei \cite{PRB97}, paying special
attention to the role played by final state interactions between the 
emitted nucleons. The present study also corrects for a mistake
in the inclusion of the $K$ and $K^*$ exchange mechanisms, which
substantially increases the ratio of neutron-induced to
proton-induced transitions, $\Gamma_n/\Gamma_p$. 
With the most up-to-date model ingredients, we find that the OME approach 
is able to describe very satisfactorily most of the
measured observables, including the ratio $\Gamma_n/\Gamma_p$.

\end{abstract}

\vspace*{0.5cm}

PACS number(s): 21.80.+a, 24.80.+y, 13.30.Eg, 13.75.Ev, 13.75.Cs

\section{Introduction}
Hypernuclei are bound systems of non-strange and strange
baryons.
In present facilities, hypernuclei are created with
hadronic reactions
--as ($\pi$,K) at Brookhaven and KEK-- or electroproduction ones --as
(e,e'K) at TJNAF--.
The decay of those objects proceeds via the weak interaction which is 10
orders of magnitude
slower than the strong one and violates parity, isospin and strangeness.
For the very light hypernuclei the mesonic decay mode ($\Lambda \to \pi N$)
is dominant, but as $A$ increases so does the Pauli blocking acting on the
outgoing nucleon, and hypernuclei mainly decay
via the one-nucleon induced non-mesonic mode, $\Lambda N \to NN$.
Since the pioneering phenomenological model of Block and Dalitz
\cite{dalitz}, many approaches have been developed to understand 
the dynamics of the decay,
and the results 
have been collected in extensive review articles
\cite{cohen,gibson,report}.
Many works are based on a meson exchange picture, either using
a simple one-pion-exchange mechanism \cite{adams,oset85}, or
including also heavier meson
exchanges, such as the $\rho$ \cite{mckellar} or the complete
pseudoscalar and vector meson octets \cite{PRB97,holstein}.
The effect of correlated-two pion exchange mechanism
\cite{itonaga,shmatikov,oset00} as well as the
role of 
$\Delta I=3/2$ transitions, implemented 
in a meson-exchange picture 
\cite{PRBM98}, have also been investigated.
A four quark weak
transition effective Hamiltonian, corrected by QCD, which contains
both $\Delta I=1/2$ and $3/2$ transitions, has also been
applied in the study of the weak decay of hypernuclei
\cite{kisslinger,oka,oka2,maltman}.
In general, the weak decay rates are reasonably reproduced by the models
but the ratio of neutron induced ($\Lambda n\to nn$) to proton
induced ($\Lambda p \to n p$) decays, $\Gamma_n/\Gamma_p$, turns out to be
smaller than the
experimental value, which is of the order of 1 or larger \cite{Sz91,No95,Mo74},
although new recent
theoretical progress has been achieved into the solution of this puzzle
\cite{oset00,oka2}.
In addition to $\Lambda N \to NN$, the decay can also proceed via the
two-nucleon induced process, $\Lambda N N\to NNN$,
originally
studied in Ref.~\cite{alberico}, which amounts to about 15\% of the total
width \cite{ramos94,ramos97,gianni} and cannot be neglected in the
experimental analysis trying to
extract the ratio $\Gamma_n/\Gamma_p$.

The high momentum of about 400 MeV/c transferred in the $\Lambda N \to NN$
reaction makes this process quite sensitive to short range physics. As a
consequence, the strong baryon-baryon interaction both in the initial and
final states plays a quite important role. 
The purpose of the present work is to revisit the one-meson-exchange
model of Ref.~\cite{PRB97}, with an especial interest in quantifying the
effect of final state interactions, as well as
the uncertainties of different model ingredients.
In doing so, we will also point out a sign error encountered
in certain transitions mediated by the exchange of strange mesons which,
when corrected, gives rise to a considerably increase in the
$\Gamma_n/\Gamma_p$ ratio. 
We will analyze the sensitivity of our results
to the way final state interactions of the
emitted nucleons are implemented, using
different prescriptions to obtain the $NN$ scattered wave function.
Our study concludes that, with the appropriate treatment of final state
interactions and with the correct sign for the contribution of the strange
mesons, the one-meson-exchange model is able to describe very
satisfactorily most of the measured observables, including the elusive
ratio $\Gamma_n/\Gamma_p$.

\section{Weak transition potential}

The weak transition potential is obtained by following the model of 
Ref. \cite{PRB97}.
In analogy to One-Boson-Exchange (OBE) based models of the strong interaction, 
the present formalism includes
not only the exchange of the long-ranged pion, 
but also more massive mesons which 
account for shorter distances. 
This potential has been presented in previous 
papers and, therefore, it is not going to be discussed here in a great detail. 
However, we would like to show its expression in coordinate space,
which represents a 
compact way of including all the mesons and transition 
channels in the mechanism.
The ${\vec r}$-space potential then reads:

\begin{equation}
V({\vec r}\,) = \sum_{i} \sum_\alpha V_\alpha^{(i)}
({\vec r}\,) = \sum_i \sum_{\alpha}
V_\alpha^{(i)} (r) \, \hat{O}_\alpha (\hat{r}) \, \hat{I}^{(i)} \, \ ,
\label{eq:compact}
\end{equation}
where the index $i$ runs over the different mesons exchanged ($i=1,\dots,
6$ represents $\pi,\eta$,K,$\rho,\omega$,K$^*$) and
$\alpha$ over the different
spin operators: central spin independent (C), central spin dependent (SS), tensor (T)
and parity violating (PV). The angular dependence is represented by the 
${\hat O}_\alpha (\hat{r}) $ operator, explicitly given by:

\[
{\hat O}_\alpha (\hat{r}) \, = \left\{
\begin{array}{ll}
\hat{1} & \mbox{C (only for vector mesons)} \, {\rm ,} \\
{\vec \sigma}_1 {\vec \sigma}_2  & \mbox{SS} \, {\rm ,}\\
S_{12} ({\hat r}) = 3 {\vec \sigma}_1 {\hat r}{\vec \sigma}_2 {\hat r} -
{\vec \sigma}_1 {\vec \sigma_2} & \mbox{T} \, {\rm ,}\\
{\rm i} \,\, {\vec \sigma}_2 {\hat r} & \mbox{PV (for pseudoscalar mesons)} \, {\rm ,} \\
\left[ {\vec \sigma}_1 \times {\vec \sigma}_2 \right] {\hat r}& \mbox{PV
(for vector mesons)}  \, {\rm ,} \\
\end{array}
\right.
\]
while the isospin operator, $\hat{I}^{(i)}$, takes the form 
${\vec \tau}_1 \, {\vec \tau}_2$ for isovector mesons ($\pi$,$\rho$), 
${\hat 1}$ for isoscalar mesons ($\eta$,$\omega$) and a combination of both
operators for the isodoublet ($K$,$K^*$).

The different pieces of the potential, $V_\alpha^{(i)} (r)$, 
are found by Fourier
transforming the potential in momentum space. 
In particular, for pseudoscalar mesons
the ${\vec q}$-space potential reads:

\begin{equation}
{V_{ps}}^{(i)}({\vec q}\,) = - G_F m_\pi^2
\frac{ g_{\scriptscriptstyle{\rm BB M}^{(i)}} }{2 M_2} \left(
A^{(i)} + \frac{B^{(i)}}{2 M_1}
{\vec \sigma}_1 \, {\vec q} \,\right)
\frac{{\vec \sigma}_2 \, {\vec q}\, }{{\vec q}^{\; 2}+{\mu^{(i)} \,}^2} \,
\hat{I}^{(i)} \ ,
\label{eq:pion}
\end{equation}
where $G_F {m_\pi}^2=2.21 \times 10^{-7}$ is the Fermi weak constant times the 
pion mass squared,
$g_{\scriptscriptstyle{\rm BB M}^{(i)} }$ the coupling at the strong baryon-baryon-meson 
(${\rm BBM}^{(i)}$) vertex, 
$A^{(i)}$ and $B^{(i)}$ the parity violating (PV) and parity conserving (PC)
weak couplings respectively, 
${\mu^{(i)} \,}$ the meson mass, and
$M_2$ ($M_1$) the average of the baryon masses 
at the strong (weak) vertex.

We want to note here that the convention has always been to direct the momentum
towards the strong vertex and, in connection to this, care must be taken when
this expression is applied to $K$ ($K^*$) exchange, since in this case vertex 1
refers to $NNK(K^*)$ and vertex 2 to $\Lambda NK(K^*)$. Therefore, the
combination of the
strange meson transition amplitudes
with the other non-strange meson contributions requires to exchange, in the
former ones, the space, spin and isospin quantum numbers between the $\Lambda$
and $N$ in the initial state and between the two emitted nucleons in the
final state. For a particular transition from an initial
$\Lambda N(L_0,S_0,T_0)$
channel to a final $NN (L,S,T)$ one, this exchange operation 
introduces the additional
factor:

\begin{equation}
f_{K (K^*)}= (-1)^{L_0+S_0+T_0} \times (-1)^{L+S+T} \times
(-1)^{t_1 + t_2} \, (-1)^{t_N+t_\Lambda} \ ,
\label{eq:fac1}
\end{equation}
where $t_1 = t_2 = t_N = \frac{1}{2}$ is the nucleon isospin and 
$t_\Lambda$ the isospin assigned to the $\Lambda$ 
isospurion, which can take the values 
$\frac{1}{2}$ or $\frac{3}{2}$ depending on whether one is
considering a change in isospin of 1/2 or 3/2.
The way to implement $\Delta I=3/2$ transitions 
in the present mechanism
can be found in Ref. \cite{PRBM98}.
The factor in Eq. (\ref{eq:fac1}) is nothing but a sign in front of the PC and/or PV 
amplitudes, which is:

\begin{equation}
f_{K (K^*)} = \left\{
\begin{array}{ll}
{\rm PC} & \left\{ \begin{array}{ll}
 +1 & \Delta I=\frac{1}{2} \, {\rm ,} \\
    & \\
 -1 & \Delta I=\frac{3}{2} \, {\rm ,}\\
\end{array} \right.
\\
& \\
{\rm PV} & \left\{ \begin{array}{ll}
(-1)^{1+S_0+S} & \Delta I=\frac{1}{2} \, {\rm ,} \\
    & \\
(-1)^{S_0+S} & \Delta I=\frac{3}{2}  \, {\rm .}\\
\end{array} \right.
\end{array}
\right.
\label{eq:fac2}
\end{equation}

It is common to write the total decay as the sum of definite
spin-isospin amplitudes. For example, for s-shell hypernuclei one 
has\cite{dalitz}:

\begin{eqnarray}
a &:& ^1S_0 \to ^1S_0 \, {\rm ,} \nonumber \\
b &:& ^1S_0 \to ^3P_0 \, {\rm ,} \nonumber \\
c &:& ^3S_1 \to ^3S_1 \, {\rm ,} \nonumber \\
d &:& ^3S_1 \to ^3D_1 \, {\rm ,} \nonumber \\
e &:& ^3S_1 \to ^1P_1 \, {\rm ,} \nonumber \\
f &:& ^3S_1 \to ^3P_1 \, {\rm .}
\label{transitions}
\end{eqnarray}

If we  adopt the $\Delta I=\frac{1}{2}$ rule -- which we always do for pseudoscalar
meson exchange \cite{PRBM98}--, the kaon and pion potentials differ in a 
minus sign only
in the $f$  amplitude in (\ref{transitions}), the only PV amplitude
with $S=S_0 (=1)$. 

In order to account for finite size effects we include a monopole form factor
at each vertex, $F^{(i)}({\vec q\,}^2)=
({\Lambda^{(i)}}^2-{\mu^{(i)}}^2)/({\Lambda^{(i)}}^2+{\vec
q\,} ^2)$, where the value of the cut-off, $\Lambda^{(i)}$, depends on the meson
($\mu^{(i)}$). The updated expression for the regularized 
potential was given in Ref.
\cite{PRKB99}. In previous calculations we used the cut-offs given by the
J\"ulich B interaction \cite{juelich}. The reason was that unlike the
early $YN$ Nijmegen
model\cite{nij89}, which used different cut-offs depending on
the irreducible representation of the baryon-baryon channel,
the J\"ulich B was the only OBE model which used different cut-offs
depending on the meson. However, the Nijmegen group has recently made
available new
baryon-baryon interactions in the strangeness $S=0,-1,-2,-3$ and $-4$ sectors
\cite{nij99}. These potentials are based on SU(3) extensions of the models in
the $S=0$ and $-1$ sectors, which are fitted to the experimental data. The
authors of Ref.~\cite{nij99} give six different models, which fit the
available $NN$ and $YN$ scattering data equally well but are characterized by
different values of the magnetic vector $F/(F+D)$ ratio, ranging from 0.4447
(model NSC97a) to 0.3647 (model NSC97f). The advantage of these new models is
that the form factors depend on the SU(3) type of meson. The momentum space
potential for each meson is multiplied by the regularizing factor
of Gaussian type, ${\rm
exp}(-\vec{q}\,^2/\Lambda^2)$, with a cut-off $\Lambda_1$ for the singlet
meson, $\Lambda_8$ for the non-strange members of the meson octet
and  $\Lambda_K$ for the
strange meson. 
In order to accommodate to our own formalism, which uses a monopole form
factor at each vertex, we will match the Gaussian to a function of the type
$[\widetilde\Lambda^2/(\widetilde\Lambda^2 + \vec{q}\,^2)]^2$ at
$|\vec{q} \, |\simeq 400$ MeV/c, the most relevant momentum transfer in the
$\Lambda N \to NN$ process. Since, by definition, both functional forms give 1
at $\vec{q}=0$, our alternative expression with the modified cut-offs 
$\widetilde\Lambda$,
listed in Table \ref{tab:formfac}, gives an excellent
reproduction of the Gaussian NSC97 form factors up to a momentum transfer 
of about 600 MeV/c.

As it is well known, one of the sources of uncertainty in OBE models comes from the 
coupling constants between baryons and mesons. 
In the strong sector the different interaction models use SU(3) 
in order to obtain the ${\rm BBM}^{(i)}$ couplings that are not 
constrained experimentally.
In the weak sector, 
only the decay of the $\Lambda$ and $\Sigma$ hyperons into nucleons and pions can be
experimentally observed. For the other mesons,
$SU_{w}(6)$ represents a convenient tool to obtain the 
PV amplitudes, while for the PC ones, we use a pole
model\cite{holstein} with only 
baryon pole resonances. Details of how these coupling constants are 
derived can be
found in Ref.~\cite{PRB97} and, in the
recent work on the weak decay of double-$\Lambda$ hypernuclei \cite{PBR00}, one
can find an updated list of the S-wave (PV) and P-wave (PC) coupling constants, the latter
obtained from the pole model using different parametrizations of the strong BB
interaction.

A measure of the amount of parity-violation in the weak decay, is given by the 
asymmetry in the angular distribution of protons coming from the decay of
polarized hypernuclei. This asymmetry is given by:
\begin{equation}
{\cal A} = P_y \displaystyle\frac{3}{J+1} \, \displaystyle\frac{Tr ({\cal M} S_y
{\cal M}^\dagger)}{Tr({\cal M} {\cal M}^\dagger)} \equiv P_y A_p \ ,
\label{asym}
\end{equation}
where $P_y$ is the polarization of the hypernucleus, characteristic of the 
production reaction, and $A_p$ the hypernuclear asymmetry parameter, 
characteristic of the weak decay.
In the expression above, $J$ is the spin of the hypernucleus, 
$S_y$ is the $J$-spin operator along the direction
perpendicular to the reaction plane, and ${\cal M}$ the hypernuclear transition 
amplitude. In order to compare with experiments, one has to multiply 
$A_p$ by the model dependent quantity
$P_y$, which has to be determined theoretically for each 
hypernucleus\cite{IM94}.
By using a shell-model for the initial hypernucleus and
assuming spherical configuration with
no mixing, one can express this amplitude in terms of two-body transitions, 
$\Lambda N \to NN$. The dependence of weak decay observables on the 
deformation of the initial (p-shell) hypernucleus was investigated 
in Ref. \cite{HP00}, 
by means of the Nilsson model with angular momentum projection.
It was found that deformation effects change these observables by about 10\%
from the spherical limit, a deviation that although non-negligible it is
smaller than the present experimental uncertainties. 

\section{Effects of the strong interaction}

Because of the lack of stable hyperon
beams, access to the
$\Delta S=-1$ $\Lambda N$ interaction is limited right now to the decay
of hypernuclei. Hence, extracting information of the elementary weak two-body
interaction requires a careful investigation of the many-body nuclear
effects present in the hypernucleus.

On the one hand, one must consider that the interacting nucleon and
$\Lambda$ hyperon are bound in the nucleus and they should be described by
bound-state single-particle wave functions, obtained from
appropriate mean-field or Hartree-Fock potentials. Note, however, that
since the mass excess
of 176 MeV in the initial state is converted into kinetic energy of the
final particles, the nucleons emerge with a large momentum of about $400$
MeV/c
and the decay is not very sensitive to the details of the single-particle
wave function. In Ref.~\cite{gianni} it is shown that the decay rates
obtained from
various realistic $\Lambda$ wave functions differ by at most 15\%.
On the other hand, the large momentum transfer implies that the $\Lambda
N\to NN$ decay process is very sensitive to the short-range correlations
induced by the strong interaction.
In the initial system, one must then replace the mean-field two-particle
$\Lambda N$ wave function by a correlated one that accounts for
the effects of the strong $YN$ interaction at short distances, 
which are not considered
in mean-field models.
Correlated wave functions,
obtained from the soft and hard core Nijmegen $YN$ interactions
\cite{nij89,hard}
by solving the corresponding finite nucleus scattering amplitude
($G$-matrix) \cite{halderson}, were compared
in Ref.~\cite{sitges}.  The differences between the wave functions
obtained with the two models are already significant
below 0.75 fm and give rise to decay
rates that differ by slightly more than 15\%. However,
since the present $YN$ models are not constrained enough by the
scattering data to resolve this discrepancy, we will admit this
uncertainty in our initial $\Lambda N$ wave function. This
uncertainty also justifies the use of
a spin-independent parametrization for the $\Lambda N$ correlation
function which, when multiplied with the uncorrelated one, was shown to
give a decay rate in between those using the soft and hard core
correlated wave functions \cite{sitges}.

Any realistic calculation must also take into account the fact that the
two nucleons
emerging from the decay feel their mutual influence, as well as that from
the residual (A-2)-particle system. However, as  mentioned before, the
most important
contribution to the decay comes from the so-called 
{\sl back-to-back} kinematical situation,
in which the two nucleons emerge with the largest possible momentum of
around 400 MeV/c.
For these fast moving nucleons the distortions with the residual
nucleus should be small, and the importance of such effects are further
diminished for inclusive observables such as the decay rates. In other 
words, one could take into account the interaction of the emitted
nucleons with the residual nucleus through an optical potential. However, 
the real part should play a minor role at those high energies, and the
imaginary part will remove flux from the $NN$ channel which will reappear
in other multi-nucleon channels, such that the total strength or decay
rate, which is the quantity we are interested in, will not be modified.
We would face a completely different situation if we were 
interested in
calculating the energy distribution of the nucleons, in which case one
should consider these final state interaction effects, which 
can change the energy, direction and charge of the primary nucleons
emitted in the weak $\Lambda N \to NN$ transition, producing as well
low energy secondary nucleons.
The Monte Carlo simulation of Ref.~\cite{ramos97} finds indeed that the
final state interactions affect the nucleon distributions mostly at
energies below 50 MeV. In the particular case of $^{12}_\Lambda$C, it is
found that final state interactions produce 
roughly 1/3 more protons at those low energies, half of which
are due to charge-exchange reactions, but barely affect the distribution at
higher energies and leave the decay rate intact.

Since in the present work we will only be interested in the decay rates,
we will limit the treatment of final state interactions to those
related to the mutual influence between the two emitted nucleons.
In contrast to the strong $\Lambda N$ interaction, the nucleon-nucleon
one is much better constrained by the huge amount of scattering data and,
although  differences in the wave function can also be observed with
various potential models,
they are
only significant below 0.5 fm,
having a moderate influence in the decay rates, as we will see.

In the literature, one finds a variety of ways of dealing with the FSI between
the emitted nucleons. Some works do not include FSI,
others use a phenomenological correlation function that simply multiplies the
uncorrelated one, and others use various approximations to the
$NN$ scattering equation. In the next section we will analyze the
sensitivity of the non-mesonic decay observables to these various choices
of FSI. The details on how
the scattering $NN$ wave function is obtained from the 
Lippmann-Schwinger ($T$-matrix) equation, as well as from other
simplified approximations, are given in the appendix.

\section{Results and discussion}

We start this section by showing the weak decay observables for
\hyp{5}{He} and \hyp{12}{C} obtained
with the model parameters used in Ref.~\cite{PRB97},
but correcting the sign in the appropriate $K$ and $K^*$ amplitudes, as pointed
out in Sect. II. 
Hence, we use the coupling constants of the NSC89
strong $YN$ interaction \cite{nij89}, together with a monopole form factor of
the type $F(\vec{q}\,^2)= (\Lambda^2-\mu^2)/(\Lambda^2+\vec{q}\,^2)$ at each
vertex. Each meson has a different cut-off value, $\Lambda$, and, since this
particular
Nijmegen model distinguishes cut-offs only in terms of the transition channel, we
choose, as in Ref.~\cite{PRB97}, the values of the J\"ulich $YN$ interaction
\cite{juelich}. The results are collected in Table \ref{tab:paper1} and we remind here 
the main results given in Ref.~\cite{PRB97} for \hyp{12}{C}, namely a non-mesonic rate of
$\Gamma_{nm}=0.75\,\Gamma_\Lambda$, where $\Gamma_\Lambda=3.8\times$ 10$^9$ s$^{-1}$ is
the decay rate of a free $\Lambda$, and a neutron-to-proton ratio
$\Gamma_n/\Gamma_p=0.068$.
 
As we see, the corrected model gives a slightly larger decay rate, 
together with a remarkable
increase in the neutron-to-proton ratio. This is due to the fact that the corrected sign
gives rise to a constructive interference between the $\pi$ and $K$ contributions to the
$I=1$ $^3S_1 \to ^3P_1$ PV amplitude, instead of the destructive one found in
Ref.~\cite{PRB97} and displayed in Fig.~8 of that work.  On the other hand, the
incorporation of $K$ exchange to the $\pi$ mechanism lowers the PC amplitudes. This was
pointed out in Fig.~7 of Ref.~\cite{PRB97} and Figs.~4 and 5 of Ref.~\cite{hypertriton},
where the contributions of all different mesons are displayed, the $\pi$ and $K$ ones
showing a destructive interference in the PC amplitudes.  
The present constructive
interference in the PV channels is especially relevant in the neutron-induced rate,
$\Gamma_n$, which lacks from the PC tensor transition for $L=0$ initial states.
Conversely,
the proton-induced rate, $\Gamma_p$, is dominated by the
PC tensor amplitude, hence its value is lowered when $K$
exchange is added to the $\pi$ mechanism, although the reduction will now be more
moderate due to the constructive interference in the less dominant PV transition.
Altogether, the correct incorporation of $K$ exchange produces, in comparison to the
results given in Ref.~\cite{PRB97}, a
more moderate decrease in the total non-mesonic decay rate, hence a slightly
larger rate is now obtained, and a \gngp ratio which is almost a factor of 3
larger.
We note that these effects of $K$ exchange have also been pointed out in
Ref.~\cite{oka2}, 
where the $\pi$ and $K$ mechanisms are considered together with 
a description of the decay in terms of quark degrees of freedom,
and in the recent work of Ref.~\cite{oset00}, where, in addition of
$\pi$ and $K$ exchange, the role of correlated two-pion exchange was also studied.  

Results for the asymmetry of the emitted protons from the weak decay of polarized
hypernuclei are also listed in Table \ref{tab:paper1}. Since this is an 
observable tied to
the interference between PV and PC amplitudes, it will also be influenced 
by the change of sign in the abovementioned PV $K$ and $K^*$
transitions. Indeed, the magnitude of the asymmetry $A_p$ has
increased almost a factor of 2 from the values given in Ref.~\cite{PRB97} 
which are $-0.27$
for \hyp{5}{He} and $0.16$ for \hyp{12}{C}.

The fact that the new Nijmegen potentials \cite{nij99} include
different
form-factors depending on the meson, makes it possible to treat
all the strong interaction ingredients involved in the weak decay transition $\Lambda N
\to NN$ consistently within the same model. We remind that the strong
interaction plays a role not only in the
strong vertices and form factors, but also in the PC piece of the weak vertex
through the pole model and in the determination  of the distorted final state
$NN$ wave function. In Table \ref{tab:paper2} we show
the decay observables for \hyp{5}{He} and \hyp{12}{C} obtained with the models
NSC97a and NSC97f.
 
\begin{figure}[hbt]
\begin{center}
\includegraphics[width=12cm,angle=-90]{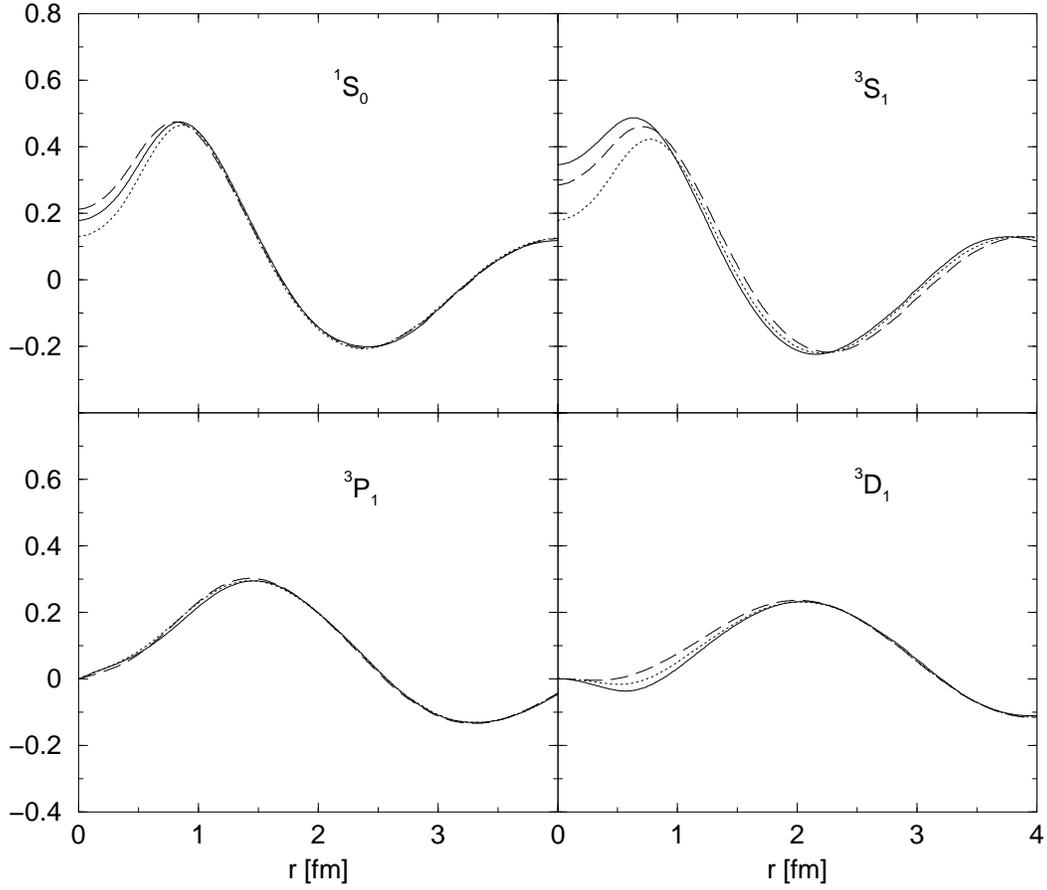}
\caption{NN wave function in various channels for a relative momentum $p_r=386$ MeV/c,
obtained from different models of the NN interaction: NSC97f \protect\cite{nij99} (solid
line),
NSC93 \protect\cite{nij93} (dotted line) and Bonn B \protect\cite{bonn} (dashed
line).
To avoid line overcrowding, only
the real part of the wave function is shown in the plots.}
\label{fig:pots}
\end{center}
\end{figure}

We observe substantial differences between both models. These might
come from the different coupling constants used in the strong and weak vertices
of the transition $\Lambda N \to NN$ amplitude, or from the different distorted
wave functions. To disentangle both effects, we show in Table \ref{tab:paper3}
the results for $^5_\Lambda$He, obtained using the same strong 
interaction model for the vertices,
which we choose to be the NSC97f one, and different distorted NN wave
functions obtained with
four NN interaction models, namely the NSC97a and f \cite{nij99}, the NSC93
\cite{nij93} and the Bonn B potential \cite{bonn}. The resulting
NN wave functions in various relevant
channels are shown in Fig.~\ref{fig:pots} for a relative momentum $p_r=386$ MeV/c.
Some minor differences  appear only in the short distances, which should affect the
decay rates moderately. By inspecting the results of
Table \ref{tab:paper3}, one can indeed see that the non-mesonic decay rates in \hyp{5}{He}
vary from 0.3 to 0.4\,$\Gamma_\Lambda$, the neutron-to-proton ratio \gngp from 0.45 to
0.5, and the asymmetry from $-0.7$ to $-0.6$.
In conclusion, it is mainly the differences
associated to the strong couplings and their influence on the weak ones the
origin for the differences observed in Table \ref{tab:paper2}. 
Going back to this table, we see that there are already some differences
between the results obtained with models NSC97a and NSC97f when the $K$
meson contribution is added to the $\pi$ meson one, but this cannot explain the
differences in the final results 
when all mesons are included. Compare, for instance, the nonmesonic
decay rate in \hyp{5}{He} of 0.43 obtained with the NSC97a model versus the value 0.32
obtained with the NSC97f one. In fact, 
by analyzing the weak couplings listed in
Ref.~\cite{PBR00}, we can see some appreciable differences for the $NNK$
coupling constants, which are larger in the NSC97f model, thus enhancing the
destructive interference of the $K$ meson and producing a reduced rate in this case.
We also observe that 
the weak vector PC $NN\omega$ coupling constant obtained from the NSC97a model is
one order of magnitude larger than that obtained from the NSC97f one. This is a consequence
of a subtle cancellation in the pole model expression, which involves the
difference between
the strong $NN\omega$ and $\Lambda\Lambda\omega$ coupling constants. Although both models
use the same strong $NN\omega$ couplings, the differences in the $\Lambda\Lambda\omega$
ones induced
by the  freedom allowed in the magnetic $F/(F+D)$ ratio may give rise, as is the case
here, to substantial differences in the values of the PC weak coupling constants. 
The larger value found in the case of the NSC97a model enhances the
constructive interference   
of the $\omega$ meson contribution and explains the final difference between the decay
rate obtained with the two Nijmegen models.
This discussion makes also clear that one must
admit this amount of uncertainty on meson-exchange models describing the nonmesonic weak
decay of hypernuclei.
We should therefore conclude that
the nonmesonic decay rate for \hyp{12}{C} ranges between 0.55 and 0.73\,$\Gamma_\Lambda$,
the neutron to
proton ratio \gngp between 0.29 and 0.34, and the hypernuclear asymmetry, 
more stable, is
0.36. In \hyp{5}{He}, the rate obtained lies in the range
0.32---0.43\,$\Gamma_\Lambda$, \gngp
in the range 0.34---0.46 and the hypernuclear asymmetry parameter is $-0.68$.
We observe that our results for the nonmesonic decay rates compare satisfactorily with
the experimental data, especially after considering that the size of the nonmesonic
two-nucleon induced mechanism, not included here, is around 0.3\,$\Gamma_\Lambda$ for
medium to heavy hypernuclei \cite{ramos94,gianni}. The ratio \gngp is considerably
larger than that found in our previous works and it now lies  practically within
the lower side of the experimental errors. We note that a larger ratio could
have been obtained, as we show below, if we had used a phenomenological treatment
of FSI effects or we had ignored them althogether. However, we want to stress
here that the
results shown in Table \ref{tab:paper2} have been obtained consistently within
the same strong interaction model and that involves also the
dynamical generation of the corresponding $NN$
wave function through a Lippmann-Schwinger equation.

The asymmetry parameter $A_p$ has increased by more than a factor of two with respect
to our previous works. As it has already been mentioned in connection to the
results
of Table~\ref{tab:paper1},  this is a direct consequence of the change of sign in the
$K$ and $K^*$ exchange PV amplitudes for transitions that do not change the spin $S$,
such as $f$ in Eq.~(\ref{transitions}). Indeed, this amplitude interferes
with $c$ and $d$ and the magnitude of the latter tensor amplitude is of
especial
relevance in meson-exchange models, as the one employed here. The increased
asymmetry
values compare less favourably with the experimental data. While the result
for \hyp{12}{C} is still within errors, the large negative value obtained here for
\hyp{5}{He} is very far away from the positive small result obtained in
Ref.~\cite{Aj00}. We note that, using a weak coupling scheme, one can relate the
hypernuclear asymmetry parameter $A_p$ to that associated to the elementary reaction
$\vec{\Lambda} N \to NN$, $a_\Lambda$ (see Eq.~(9) in Ref.~\cite{PRB97}). In
the case of
\hyp{5}{He}, $a_\Lambda=A_p$ and for \hyp{12}{C} the relation is $a_\Lambda=-2A_p$.
Therefore the parameter $a_\Lambda$ of the elementary reaction that can be extracted
from our results in Table \ref{tab:paper2}, turns out to be very similar in
both hypernuclei. However, the present experimental results seem to
contradict this finding because a small positive value is obtained for
\hyp{5}{He} \cite{Aj00}, while a large negative one was found in
\hyp{12}{C} \cite{Aj92}. In
Ref.~\cite{Aj00} arguments are
given to try to understand these differences on the basis
of the fact that \hyp{5}{He} is a s-shell nucleus and
\hyp{12}{C} a p-shell one, the latter one allowing transitions from initial
$\Lambda N$ relative $P$-wave
sates. However, it was already noticed that most of
the decay for p-shell nuclei comes
from relative $S$-wave initial
states \cite{bennhold92} and we have checked here that removing the initial $P$-wave
amplitudes, the decay rate of \hyp{12}{C} using the NSC97f model gets slightly
reduced to
$0.48$ and the asymmetry parameter $A_p$ only changes to $0.34$.
Hence, meson-exchange models do not explain the present experimental differences
between the
elementary asymmetry for \hyp{5}{He} and \hyp{12}{C}, and this seems to be a
new puzzle in the study of the weak decay of hypernuclei. This poses a
challenge to
the theoretical models and calls for further experimental efforts. Besides valuable new
information from hypernuclear weak decay experiments,
it is also possible to gain direct information from the inverse reaction $pn\to p
\Lambda$, already studied theoretically \cite{PRKB99,haiden95,kishi99,inoue00} and
presently
under preparation at RCNP (Osaka).

\begin{figure}[hbt]
\begin{center}
\includegraphics[width=12cm,angle=-90]{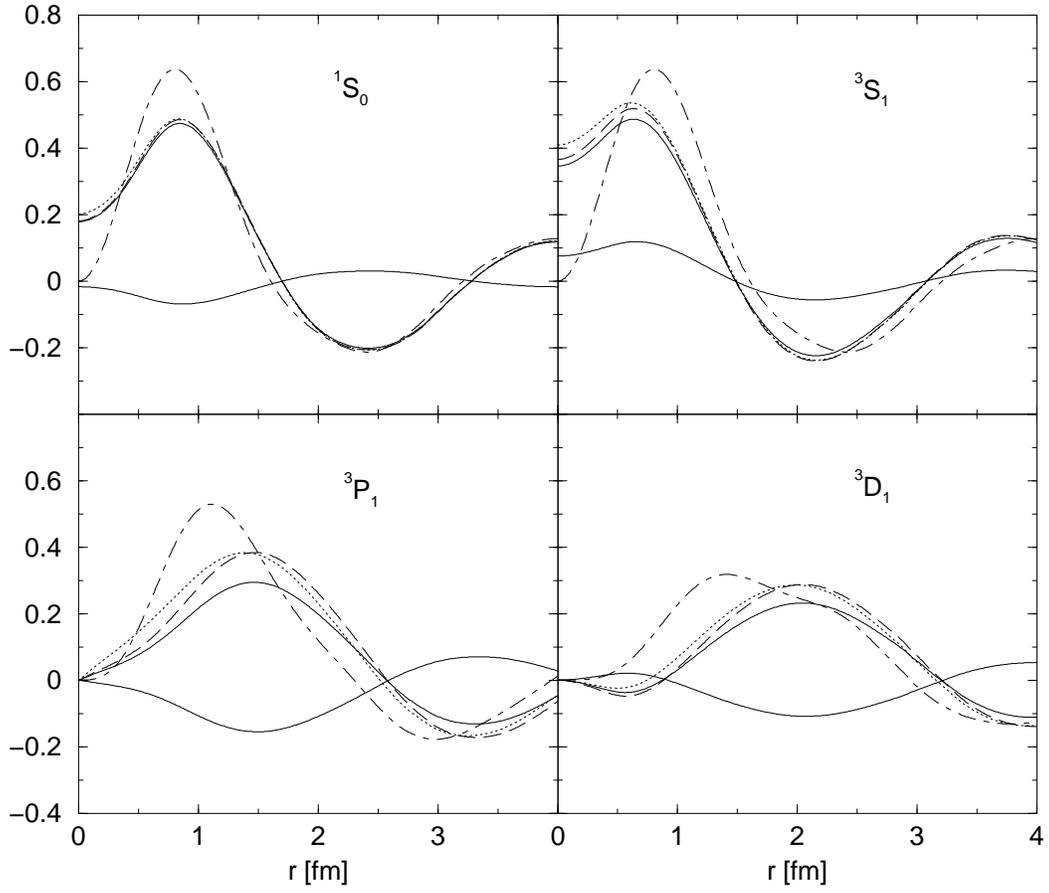}
\caption{NN wave function in various channels  for a relative momentum $p_r=386$
MeV/c, obtained from
different treatments in the solution of the scattering equation:         
$T$-matrix (solid lines), Real part of $T$-matrix (dotted line),
$K$-matrix (dashed line) and phenomenological (dot-dashed line).
}
\label{fig:fsiall}
\end{center}
\end{figure}

One of the purposes of this paper was to study the influence of FSI in the 
$\Lambda N \to NN$ transition. The results of Table \ref{tab:paper3} have already shown
that, using
NN wave functions obtained from various realistic NN potentials, one obtains similar
values for the weak decay
observables. However, one finds in the literature several possibilitites for the treatment
of FSI. They are either ignored, or included through a
phenomenological function, such as $f_{\rm phen}=1-j_0(q_cr)$ with $q_c=3.93$
fm$^{-1}$, multiplying the  
free NN plane wave, or through a distorted wave function built up from the 
$K$-matrix  rather than from the $T$-matrix.
In Fig.~\ref{fig:fsiall} we show the resulting NN wave functions in various channels
for a relative momentum $p_r=386$ MeV/c, obtained from the various approaches to FSI. The
solid lines represent the real and imaginary parts of the
wave function obtained from the
proper solution of the Lippmann-Schwinger equation ($T$-matrix). The imaginary part
is the minoritary component in all panels. The dotted line represents the
result
obtained when only the real part of the $T$-matrix is retained in the construction of the
distorted wave function from Eq.~(\ref{eq:psipw}), which is further approximated by its
principal part value. This was the prescription followed in Ref.~\cite{PRB97}.
The dashed line corresponds to a $K$-matrix approximation, which would be the appropriate
solution for standing waves, i.e. non-propagating solutions, as is the case in the nuclear
medium, where Pauli blocking prevents two interacting nucleons of the Fermi sea
from exciting intermediate NN states that are already occupied by other nucleons. Finally,
the dot-dashed line displays the product of the phenomenological correlation function
with the corresponding non-interacting Bessel function.
We observe important differences beween the various approaches to the treatment of FSI,
especially in the case of a phenomenological correlation function. Obviously, this will
have an influence on the decay observables. As an example, we show in Table
\ref{tab:paper4} the results for \hyp{5}{He}, obtained using
the NSC97f model and taking all these different treatments for
the distorted NN wave functions.
The rates can differ substantially, by more than a factor of two in some
cases. The neutron-to-proton ratio can also be fictitiously enhanced by using the
phenomenological approach to FSI effects, or no FSI at all, while the asymmetry
remains more stable.

\section{Conclusions}

We have revisited the OME model of  Ref.~\cite{PRB97} for the weak 
nonmesonic decay of hypernuclei to the light of the new Nijmegen 
baryon-baryon potentials \cite{nij99}. These strong interaction models
influence the weak decay mechanism, not only through the coupling constants and
form factors at the strong vertex involved in the $\Lambda N\to
NN$ reaction, but also via the PC piece of the weak vertex, obtained from a pole model,
as well as from the corresponding correlated wave functions for the initial $\Lambda N$
and final $NN$ states.

We have paid an especial attention to disentangle the influence of the various model
ingredients involved in the decay mechanism. 
In particular, we have found that the
uncertainties tied to the strong coupling constants are a source of uncertainty on the
nonmesonic weak decay observables. Employing different strong models of the Nijmegen
group, and working consistently within each one, we estimate the decay rate of
\hyp{5}{He} to be in the range 0.32---0.43\,$\Gamma_\Lambda$, 
the neutron-to-proton ratio
in the range 0.34---0.46 and the hypernuclear asymmetry parameter $A_p$ around
$-0.68$. In the case of
\hyp{12}{C}, the decay rate turns out to be in the range
0.55---0.73\,$\Gamma_\Lambda$, the neutron-to-proton ratio in the range
0.29---0.34 and the asymmetry $A_p$  around $0.36$.
The new results for the nonmesonic rates compare favourably with the present
experimental
data.
The ratio \gngp has increased with respect to our previous works and it now
lies
practically within the lower side of the error band. The asymmetry for \hyp{12}{C} is
also compatible with experiment \cite{Aj92} but that for \hyp{5}{He} disagrees strongly from
the recent experimental observation \cite{Aj00}. The latter work finds a small 
and positive value for the elementary asymmetry parameter $a_\Lambda$ 
in \hyp{5}{He}, while that for 
\hyp{12}{C} is large and negative.
Our meson-exchange model does not explain the present experimental differences and
understanding this issue is one of the current challenges, both experimental
and theoretical,
in the study of the weak decay of hypernuclei.

We have found a tremendous influence on the weak decay
observables from the way FSI are
considered, especially in the
case of total and partial decay rates. A phenomenological
implementation of FSI effects, or not including them at all, gives rise to decay rates
that differ by more than a factor of two, and to a neutron-to-proton ratio about 20\%
larger from what is obtained with the more realistic calculation that uses the
proper $NN$ scattering wave function.
The differences observed in the decay rates and the \gngp ratio are much larger than
the uncertainties tied to the different strong interaction 
models commented above. Therefore, accurate calculations of the
nonmesonic weak decay of hypernuclei
demand a proper treatment of FSI effects through the solution of a $T$-matrix using
realistic NN interactions.

\section*{Acknowledgments}

The authors are indebted to K. Sasaki and
M. Oka for many clarifying discussions about the inclusion of the $K$-exchange 
mechanism. We also thank C. Bennhold for many useful suggestions and 
E. Oset for discussions on the effects of an
optical nucleon-nucleus potential.
This work has been partially supported by the U.S. Dept. of Energy under
Grant No. DE-FG03-00-ER41132, by the DGICYT (Spain) under contract
PB98-1247, by the Generalitat de Catalunya project SGR2000-24, and by the
EEC-TMR Program EURODAPHNE under contract CT98-0169.     

\renewcommand{\theequation}{\Alph{section}.\arabic{equation}} 
\setcounter{section}{1}
\setcounter{equation}{1}                                         
\section*{Appendix}
The wave function describing the relative motion of two particles moving
under the influence of a two-body potential $V$ is obtained from the
Lippmann-Schwinger equation:

\begin{equation}
\mid \Psi^{(\pm)} \rangle = \mid \Phi \rangle +
\frac{1}{E-H_0 \pm {\rm i} \eta} V \mid \Psi^{(\pm)} \rangle \ ,
\label{eq:psiwf0}
\end{equation}
where $\mid \Phi \rangle \equiv \mid \vec{k} \rangle$ is the free relative
motion state,
solution of the
free Schr\"odinger equation 
with energy $E$, $H_0 \mid \Phi \rangle = E \mid \Phi \rangle$, with 
$H_0$ being
the relative kinetic energy
operator. The positive (negative) sign in front of the infinitesimal
quantity ${\rm i} \eta$ arises from the requirement that the relative
motion
is free in the remote past (future).
Alternatively, one can write the Lippmann-Schwinger equation in terms of
the transition matrix $T$, defined such that $T \mid \Phi \rangle = V
\mid \Psi^{(+)} \rangle$ (\,$\langle \Psi^{(-)} \mid V = \langle \Phi \mid
T$\,), with the result

\begin{eqnarray}
\mid \Psi^{(+)} \rangle &=& \mid \Phi \rangle +
\frac{1}{E-H_0 + {\rm i} \eta} T \mid \Phi \rangle  \, {\rm ,}
\label{eq:psip}\\
\langle \Psi^{(-)} \mid &=& \langle \Phi \mid +
\,\langle \Phi \mid T \frac{1}{E-H_0+{\rm i}\eta} \, {\rm ,}
\label{eq:psim}
\end{eqnarray}
where the $T$ operator obeys:
\begin{equation}
T = V + V \frac{1}{E-H_0 + {\rm i} \eta} T \ .
\label{eq:teq}
\end{equation}

Let's take for instance $\Psi^{(-)}$ (representing the two nucleons in the final
state) and project 
Eq. (\ref{eq:psim}) into coordinate space.
Inserting a complete set of states in the r.h.s. of Eq. (\ref{eq:psim})
and considering also the spin quantum numbers via two-particle
coupled spin states, we find:

\begin{eqnarray}
&&\langle \Psi^{(-)}_{\vec k}, S\,M_S \mid {\vec r} \, \rangle =
\langle
 {\vec k}, S\,M_S \mid {\vec r} \, \rangle \nonumber \\
&&\phantom{\Psi^{(-)}\,\,}
+
\sum_{S^\prime\,
M_{S^\prime}}
\int d^3 k^\prime\frac{  \langle {\vec k}, S\,M_S \mid T \mid {\vec
k^\prime},
S^\prime\, M_S^\prime \rangle  
\langle  {\vec k^\prime}, S^\prime\,M_S^\prime \mid  {\vec r}\,
\rangle }
{E(\vec{k}\,)-E({\vec k^\prime}\,)+{\rm i}\eta} \ ,
\label{eq:wftmat}
\end{eqnarray}
where $\langle \vec{k}, S\,M_S \mid \vec{r} \, \rangle$ stands for the
adjoint of the free plane wave,
$ \displaystyle\frac{{\rm e}^{-{\rm i} \vec{k}\,\vec{r}}}{(2\pi)^{3/2}}
\,\, \langle \, S M_S \mid$.
We perform a partial wave decomposition of the 
wave functions $\langle \Psi^{(-)}_{\vec k}, S M_S \mid
{\vec r} \, \rangle$ and \mbox{$\langle \vec{k}, S\, M_S \mid {\vec r}\,
\rangle$}, working in the coupled $(LS)J$ representation, and,
similarly to Eq. (18.119b) of Ref.~\cite{joachain}, we obtain:

\begin{eqnarray}
\langle \, \Psi^{(-)}_{\vec k}, S\, M_S \mid {\vec r} \, \rangle &=&
\sqrt{\frac{2}{\pi}} \sum_{L L^\prime S^\prime J M} (- {\rm i})^{L^\prime}
\, 
\Psi^{(-) \, *\, J}_{L^\prime S^\prime, L S} (k,r) \nonumber \\
&\times& \sum_{M_L}  \langle L M_L S M_S | J M \rangle \, Y_{L M_L} ({\hat k})
\, {\cal J}^{M\, \dagger}_{L^\prime S^\prime J}(\hat{r}) \ ,
\label{eq:pwwf} 
\end{eqnarray}
where ${\cal J}^{M\, \dagger}_{L^\prime S^\prime J}(\hat{r})$ is the
adjoint of the
generalized spherical harmonic given by
\begin{equation}
{\cal J}^{M}_{L^\prime S^\prime J} (\hat{r}) = \sum_{M^\prime_L
M^\prime_S} \langle J M \mid
L^\prime M_L^\prime S^\prime M^\prime_S \rangle \, Y_{L^\prime
M^\prime_L} ({\hat r}) \, \mid S^\prime M^\prime_S\rangle \ .
\label{eq:spher}
\end{equation}
The partial wave decomposition for the free wave function  is
simply obtained by replacing in
Eq.~(\ref{eq:pwwf})
\begin{equation}
\Psi^{(-)\, *\, J}_{L^\prime S^\prime, L S} (k,r) 
\to j_L (kr) \delta_{L L^\prime} \delta_{S S^\prime} \ ,
\label{eq:pwfree}
\end{equation}
where $j_L(kr)$ is the spherical Bessel function.

A similar decomposition may be written down for the $T$-matrix elements,
\begin{eqnarray}
\langle \vec{k} , S\,M_S \mid T \mid {\vec k^\prime}, S^\prime\,
M^\prime_S \rangle &=& \sum_{J M} \sum_{L M_L} \sum _{L^\prime M^\prime_L}
Y_{L M_L}(\hat{k}) Y^*_{L^\prime M^\prime_L}(\hat{k^\prime}) \nonumber \\
& \times & 
\langle L M_L S M_S \mid J M \rangle
\langle L^\prime M^\prime_L S^\prime M^\prime_S \mid J M \rangle
\nonumber \\
& \times & \langle k (LS) J M \mid T \mid k^\prime (L^\prime S^\prime) J M
\rangle \ .
\label{eq:pwt}
\end{eqnarray}
Inserting Eqs.~(\ref{eq:pwwf}), and (\ref{eq:pwt}) into Eq.~(\ref{eq:wftmat}),
using Eqs.~(\ref{eq:spher}) and (\ref{eq:pwfree}), and
carrying out the angular integration, one
obtains the following equation for the partial wave components of the
correlated
wave function
\begin{eqnarray}
\Psi^{(-)\, *\, J}_{L^\prime S^\prime, L S} (k,r) &=& j_L(kr) 
\delta_{L L^\prime} \delta_{S S^\prime}  \nonumber \\
& + & \int k^{\prime\,2} dk^\prime \frac{
\langle k (LS) J M \mid T \mid k^\prime (L^\prime S^\prime) J M
\rangle j_{L^\prime}(k^\prime r)}
{E(k)-E(k^\prime)+{\rm i}\eta} \ ,
\label{eq:psipw}
\end{eqnarray}
where the partial wave $T$-matrix elements fulfil the integral equation:
\begin{eqnarray}
&&\langle k (LS) J M \mid T \mid k^\prime (L^\prime S^\prime) J M  
\rangle =   \langle k (LS) J M \mid V \mid k^\prime (L^\prime S^\prime)
J M  \rangle \nonumber \\
&& ~~~~~~~~~~~~~~~~~~~~~ + \sum_{S^{\prime\prime} L^{\prime\prime}} \int
k^{{\prime\prime}\,2} 
dk^{\prime\prime} \frac{
\langle k (LS) J M \mid V \mid k^{\prime\prime} (L^{\prime\prime} 
S^{\prime\prime}) J M  \rangle} {E(k) - E(k^{\prime\prime}) + {\rm i}\eta}
\nonumber \\ 
&&~~~~~~~~~~~~~~~~~~~~~~~~~~~~~ \times~
\langle k^{\prime\prime} (L^{\prime\prime}
S^{\prime\prime}) J M \mid T \mid k^\prime (L^\prime S^\prime)
J M  \rangle \ .
\label{eq:tmatpw}
\end{eqnarray}
Both equations (\ref{eq:psipw}) and (\ref{eq:tmatpw}) are solved in
momentum space following a numerical matrix
inversion method described in Ref.~\cite{HT70}.
Isospin is easy to incorporate at the final
step of our calculation by multiplying the resulting correlated wave
function by the isospinor $\chi_T^{M_T}$, allowing only for those cases
that fulfil the antisymmetry requirement $L+S+T={\rm odd}$.

An alternative but approximate way of implementing the $NN$ final state
interactions is through the $K$-matrix operator, which obeys:
\begin{equation}
K = V + V {\cal P}\left(\frac{1}{E-H_0}\right) K \ ,
\label{eq:keq}
\end{equation}
where ${\cal P}$ indicates the Cauchy principal value. Standing waves can
then be obtained by following the same procedure of Eqs.
(\ref{eq:wftmat}) through (\ref{eq:tmatpw}) but replacing the $T$-matrix
elements by those of the real $K$-matrix.

Finally, a simple way to include the effects of the strong interaction 
between the particles is to use a phenomenological correlation function,
$f_{\rm phen}(r)$, such that
\begin{equation}
\Psi_{L^\prime S^\prime, LS}^{(-)\, * \, J} (k, r) = f_{\rm phen}(r) j_L(k r)
\delta_{L L^\prime} \delta_{S S^\prime} \ .
\label{eq:phenom}
\end{equation}
This has been the approach followed by several authors, and functions such
as a Gaussian or of the type
$f_{\rm phen}(r)=1-j_0(q_c r)$ with $q_c=3.93$fm$^{-1}$ have been used \cite{oka2}.

\newpage

\begin{table}
\begin{tabular}{|c|c|c|c|c|c|}
$\pi$ & $\eta$ & $K$ & $\rho$ & $\omega$ & $K^*$ \\
\hline
1750 & 1750 & 1789 & 1232 & 1310 & 1649 \\
\end{tabular}
\caption{Cut-off values in MeV used in the present calculation for a FF of
the type
${\widetilde{\Lambda}} 
^2/({\widetilde{\Lambda}}^2 
+ \vec{q}\,^2)$, which
matches the exponential-type FF used in Ref.~\protect\cite{nij99}.} 
\label{tab:formfac}
\end{table}                           

\begin{table}[ht!]
\def\arraystretch{1.4}
\centering
\begin{tabular}{|l|c|c|c|c|}
$^5_\Lambda$He &
$\Gamma_{nm}$ & $\Gamma_n/\Gamma_p$ & $\Gamma_p$ & $A_p$  \\
\hline
$\pi$ & 0.438 & 0.104 & 0.397 & $-0.282$ \\
\hline
$\pi+K$ & 0.321 & 0.286 & 0.249 & $-0.484$ \\
\hline 
all &  0.496 & 0.226 & 0.405 & $-0.447$ \\ 
\hline
\hline
$^{12}_\Lambda$C &
$\Gamma_{nm}$ & $\Gamma_n/\Gamma_p$ & $\Gamma_p$  & $A_p$  \\
\hline                       
$\pi$ & 0.771 & 0.093 & 0.705 & 0.205 \\
\hline
$\pi+K$ & 0.558 & 0.210 & 0.461 & 0.305 \\
\hline  
all & 
0.834 & 0.181 & 0.706 & 0.275 \\
\end{tabular}
\caption[]{Weak decay observables for $^5_{\Lambda}$He and $^{12}_\Lambda$C 
including $\pi$, $\pi + K$ and all meson
($\pi+\eta+K+\rho+\omega+K^*$) contributions.
The total and partial non-mesonic
decay rates are in units of
$\Gamma_\Lambda = 3.8 \times 10^{9}$ $s^{-1}$.
The strong NSC93 coupling constants\cite{nij89} and J\"ulich B
cut-offs\cite{juelich} in the (monopole) FF have been used.
For the final NN wave function we used the solution of a $T$-matrix equation
--Eq. (\ref{eq:wftmat})-- with the NSC93 potential model.}
\label{tab:paper1}
\end{table}                          

\begin{table}[ht!]
\def\arraystretch{1.8}
\centering
\begin{tabular}{|l|cc|cc|cc|cc|}
$^5_\Lambda$He &
\multicolumn{2}{c|}{$\Gamma_{nm}$} & \multicolumn{2}{c|}{$\Gamma_n/\Gamma_p$} &
\multicolumn{2}{c|}{$\Gamma_p$} & \multicolumn{2}{c|}{$A_p$} \\
 & a & f & a & f & a & f & a & f \\
\hline
$\pi$ & 0.424 & 0.425  & 0.086 & 0.086 & 0.390 & 0.391 & $-0.252$ & $-0.252$ \\
$\pi+K$ & 0.272 & 0.235 & 0.288 & 0.498 & 0.211 & 0.157 & $-0.572$ & $-0.606$ \\
\hline
all &
0.425 & 0.317 & 0.343 & 0.457 & 0.317 & 0.218 & $-0.675$ & $-0.682$ \\
\hline
EXP: & \multicolumn{2}{c|}{$0.41\pm 0.14$\cite{Sz91}} &
\multicolumn{2}{c|}{$0.93\pm 0.55$\cite{Sz91}} &
\multicolumn{2}{c|}{$0.21 \pm 0.07$\cite{Sz91}} &
\multicolumn{2}{c|}{$0.24 \pm 0.22$ \cite{Aj00}}  \\
\hline
\hline
$^{12}_\Lambda$C &
\multicolumn{2}{c|}{$\Gamma_{nm}$} & \multicolumn{2}{c|}{$\Gamma_n/\Gamma_p$} &
\multicolumn{2}{c|}{$\Gamma_p$} & \multicolumn{2}{c|}{$A_p$} \\
 & a & f & a & f & a & f & a & f \\
\hline
$\pi$ & 0.762 & 0.751 & 0.078 & 0.079 & 0.707 & 0.696 & 0.169 & 0.171 \\
$\pi+K$ & 0.485 & 0.413 & 0.205 & 0.343 & 0.403 & 0.308 & 0.313 & 0.320 \\
\hline
all &
0.726 & 0.554 & 0.288 & 0.341 & 0.564 & 0.413 & 0.358 & 0.367 \\
\hline
EXP: & \multicolumn{2}{c|}{$1.14\pm 0.08$\cite{Bhang98}} &
\multicolumn{2}{c|}{$1.33^{+1.12}_{-0.81}$\cite{Sz91}} &
\multicolumn{2}{c|}{$0.31^{+0.18}_{-0.11}$\cite{No95}} &
\multicolumn{2}{c|}{$0.05\pm 0.53$\footnote{This number has been
obtained dividing the experimental asymmetry,
${\cal A}= -0.01\pm 0.10$\cite{Aj92}, by a polarization of
$P_y=-0.19$\cite{RM92}.}} \\
 & \multicolumn{2}{c|}{$0.89 \pm 0.15 \pm 0.03$\cite{No95}} &
\multicolumn{2}{c|}{$1.87 \pm 0.59^{+0.32}_{-1.00}$\cite{No95}} & & & & \\      & \multicolumn{2}{c|}{$1.14\pm 0.2$\cite{Sz91}} &
\multicolumn{2}{c|}{$0.70\pm 0.3$\cite{Mo74}} & & & &  \\
 & & & \multicolumn{2}{c|}{$0.52\pm 0.16$\cite{Mo74}} & & & & \\
\hline
\end{tabular}
\caption[]{Weak decay observables for $^5_{\Lambda}$He and $^{12}_\Lambda$C
including $\pi$, $\pi + K$ and all meson
($\pi+\eta+K+\rho+\omega+K^*$) contributions.
The total and partial non-mesonic
decay rates are in units of
$\Gamma_\Lambda = 3.8 \times 10^{9} s^{-1}$.
The strong NSC97a (left column)  and NSC97f (right column) coupling constants
and
cut-offs in the FF (see Table \ref{tab:formfac}) have been used
\cite{nij99}.
For the final NN wave function we used the solution of a $T$-matrix equation
--Eq. (\ref{eq:wftmat})-- with the corresponding NSC97a 
and NSC97f potential models.}
\label{tab:paper2}
\end{table}

\begin{table}[ht!]
\def\arraystretch{1.8}
\centering
\begin{tabular}{|l|c|c|c|c|}
$^5_\Lambda$He &
$\Gamma_{nm}$ & $\Gamma_n/\Gamma_p$ & $\Gamma_p$  & $A_p$  \\
\hline
NSC97a & 0.320 & 0.459 & 0.219 & $-0.680$ \\
NSC97f & 0.317 & 0.457 & 0.218 & $-0.682$ \\
NSC93 & 0.405 & 0.483 & 0.273 & $-0.568$ \\
Bonn B & 0.398 & 0.484 & 0.268 & $-0.602$ \\
\end{tabular}
\caption[]{Weak decay observables for $^5_{\Lambda}$He 
including all meson
($\pi+\eta+K+\rho+\omega+K^*$)
contributions. The total and partial
non-mesonic
decay rates are in units of
$\Gamma_\Lambda = 3.8 \times 10^{9} s^{-1}$.
The strong NSC97f coupling constants and
cut-offs in the FF (see Table \ref{tab:formfac}) have been used \cite{nij99}.
For the final NN wave function we solve a $T$-matrix equation
using different interaction models, as quoted in the table.
}
\label{tab:paper3}
\end{table}                                

\begin{table}
\def\arraystretch{1.8}
\centering
\begin{tabular}{|l|c|c|c|c|}
$^5_\Lambda$He &
$\Gamma_{nm}$ & $\Gamma_n/\Gamma_p$ & $\Gamma_p$  & $A_p$  \\
\hline
$T$ & 0.317 & 0.457 & 0.218 & $-0.682$ \\
Re($T$) & 0.490 & 0.512 & 0.324 & $-0.655$ \\
$K$ & 0.475 & 0.471 & 0.323 & $-0.650$ \\
$f_{\rm phen} (r)$ & 0.766 & 0.619 & 0.473 & $-0.671$ \\
no FSI & 0.721 & 0.614 & 0.447 & $-0.654$ \\
\end{tabular}
\caption[]{Different approaches to the $NN$ final state wave function in the
weak decay of $^5_{\Lambda}$He. The total and partial
non-mesonic
decay rates are in units of
$\Gamma_\Lambda = 3.8 \times 10^{9} s^{-1}$. 
The strong BB NSC97f model has been used. 
}
\label{tab:paper4}
\end{table}

\end{document}